\documentclass[pra,preprint,superscriptaddress,showpacs,preprintnumbers,amsmath,amssymb]{revtex4}
\usepackage{graphicx}
\usepackage{dcolumn}
\usepackage{bm}
\usepackage{color}
\usepackage{epsfig}
\newcommand{\beq}{\begin{eqnarray}}
\newcommand{\eeq}{\end{eqnarray}}
\newcommand{\al}{\alpha}
\newcommand{\om}{\omega}
\newcommand{\lam}{\lambda}
\newcommand{\lan}{\langle}
\newcommand{\ran}{\rangle}
\newcommand{\hz}{\hbar\rightarrow 0}
\newcommand{\lhz}{\lim_{\hz}}

\input xy 
\xyoption{all}

\begin{document}

\title{
Classical States and Their Quantum Correspondence
}
\author{I. Hen}%
 \email{itayhe@post.tau.ac.il}
\affiliation{%
School of Physics and Astronomy,
Tel-Aviv University, Tel-Aviv 69978,
Israel.
}%
\author{A. Kalev}
\email{amirk@techunix.technion.ac.il}
 \affiliation{Department of Physics, Technion-Israel Institute of Technology, Haifa 32000,
Israel.}

\begin{abstract}
We point out
a correspondence between classical and quantum states, by showing that
for every classical distribution over phase--space,
one can construct a corresponding quantum state, such that
in the classical limit of $\hz$ the latter converges to the former with respect to
all measurable quantities.
\end{abstract}


\maketitle
An important concept in quantum mechanics is the
correspondence principle, first invoked by Niels Bohr in 1923,
which states that quantum mechanics should behave in a classical manner
in the limit of $\hz$. In this limit, canonical operators must commute,
Heisenberg uncertainty relations should vanish and the equations of classical physics emerge.
\par
Indeed, the behavior of quantum systems in the classical limit
has become, naturally, a central issue in quantum mechanics and is still studied extensively
within every sub--discipline of physics. It has been investigated using a variety of different approaches,
a few of which are
the WKB method, Wigner functions, Fourier integral operators and
Feynman integrals (for a review see \cite{Werner}).
\par
Although considerable progress has been made throughout the years,
the mechanism through which quantum and classical mechanics are interlaced
is still not fully understood and the exact correspondence between the theories
is not yet known.
In what follows we point out
a correspondence between classical and quantum states, by showing that
for every classical distribution over phase--space,
one can construct a corresponding quantum state, such that
in the classical limit of $\hz$ the latter converges to the former with respect to
all measurable quantities. It should be noted
that $\hbar$ must not be taken naively to zero
in obtaining the classical limit.
The mathematical procedure of taking the limit will only make sense
at the level at which expectation values are concerned \cite{QC}.
\par
For the sake of simplicity, we start off by considering
states described by only one pair of canonical variables,
though a generalization to states with many degrees of freedom can be obtained
in a rather straightforward manner, which will be discussed later on.
A pure state of a classical system with one degree of freedom
is described by a point $(x_0,p_0)$ in phase space, where
$x$ and $p$ are the usual canonical variables.
A classical ``observable'' would be any real--valued function $A(x,p)$
and a classical ``measurement'' of that observable on a state $(x_0,p_0)$
can thus be given by
\beq\label{eq:Ac}
\lan A \ran_C=\int dxdp \delta(x-x_0)\delta(p-p_0)A(x,p)=A(x_0,p_0)\;.
\eeq
A classical observable has the additional property that if one constructs
another observable $f(A)$ where $f$ is a (smooth) function of $A$,
the resultant measurement of $f(A)$ on a state $(x_0,p_0)$ would be:
\beq\label{eq:fAc0}
\lan f(A) \ran_C=\int dxdp \delta(x-x_0)\delta(p-p_0) f(A(x,p))=f(A(x_0,p_0))\;.
\eeq
This is of course not true for quantum observables. Nonetheless,
we would like to show that when the classical limit is taken, (\ref{eq:fAc0})
is true for the quantum observables we will be discussing, as well.
To make the classical--quantum correspondence,
we assign to every phase--space point $(x,p)$ a unique quantum state:
\beq \label{rhoqk}
\hat{\rho}_{(q,k)} \equiv |\al \ran \lan\al|
\eeq
where
$|\al\ran$ is a coherent state with $\al\equiv q+ik$ with
$q$ and $k$ being dimensionless variables, relating to the
dimensional $x$ and $p$
by $(q,k)=\frac1{\sqrt{\hbar}}(\lam x,p/\lam)$
($\lam$ being a $\hbar$-independent ''unit fixing'' constant, which will be taken to be $1$).\\
To every classical observable $A(x,p)$, we assign a quantum (Hermitian) operator \cite{Klauder}:
\beq\label{eq:AqOp}
\hat{A}\equiv\ \frac1{\pi} \int d^2\al A(x,p)|\al\ran\lan\al|=
\frac1{\pi} \int dqdk A(x,p) \hat{\rho}_{(q,k)}\;.
\eeq
We note that the assignment presented above is not the usual (first) ``quantization''
of classical observables (\textit{e.g.}, the quantum operator
assigned to the classical position variable $x$ is different from the quantum
position operator).
The expectation value of a measurement of $\hat{A}$ on a state $\hat{\rho}_{(q_0,k_0)}$
is:
\beq\label{eq:Aq}
\lan \hat{A} \ran_Q&\equiv& Tr [\hat{\rho}_{(q_0,k_0)}\hat{A}]=
\frac1{\pi} \int d^2\al A(x,p) Tr[
|\al_{0}\ran \lan\al_{0}|
\al\ran\lan\al|]\\\nonumber
&=&\frac1{\pi} \int d^2\al A(x,p)
 |\lan\al_{0}|\al\ran|^2\;.
\eeq
In order to take the $\hz$ limit of (\ref{eq:Aq}), we switch to
the ``dimensional'' representation by expressing every $(q,k)$ pair in terms of
$(x,p)$, so for the limit, we have:
\beq\label{eq:AQtoAC}
\lhz \lan \hat{A} \ran_Q&=&\lhz \frac{1}{\pi}\int dqdk A(x,p) |\lan\al_{0}|\al\ran|^2\\\nonumber
&=&\int dxdp
A(x,p)\lhz \frac{1}{\pi \hbar}\exp[-\hbar^{-1}
\left((x-x_0)^2+(p-p_0)^2\right)]\\\nonumber
&=&\int dxdp
A(x,p)\delta(x-x_0)\delta(p-p_0)=A(x_0,p_0)\equiv \lan A \ran_C\;.
\eeq
where we have used
$\lan\al|\al^{'}\ran=\exp[-\frac{1}{2}(q-q')^2-\frac{1}{2}(k-k')^2-i(q k'-k q')]$.

However, a correspondence between the expectation values of classical and
quantum observables
is of course not enough. One must also
require that in the classical limit the following should also hold:
\beq \label{eq:General}
\lhz \lan f(\hat{A}) \ran_Q =\lan f(A) \ran_C \;.
\eeq
Supposing that $f(A)$ has the Taylor series expansion $\sum_n f_n A^n$, (\ref{eq:General})
reduces to the requirement that
\beq \label{eq:An}
\lhz \lan \hat{A}^n \ran_Q =\lan A^n \ran_C \;.
\eeq
In order to show just that, let us compute the $\hz$ limit
of the expectation value of
the $(n-1)$-th moment of the quantum observable $\hat{A}$
which is given by:
\beq\label{eq:AnQtoAnC}
\lhz \lan \hat{A}^{n-1}\ran_Q&\equiv& \lhz Tr [\hat{\rho}_{(q_0,k_0)}\hat{A}^{n-1}]
=\lhz \int \left( \prod_{i=1}^{n-1} \pi^{-1} dq_i dk_i A(x_i,p_i) \right)\\\nonumber
& \times &
\lan\al_{0}|\al_{1}\ran\lan\al_{1}|\al_{2}\ran
\cdots\lan\al_{n-1}|\al_{0}\ran\\\nonumber
&=&\int \left( \prod_{i=1}^{n-1} dx_i dp_i A(x_i,p_i) \right)
\lhz \frac{\exp[-\hbar^{-1}\mathbf{u}^{\dag} V\mathbf{u}]}{\hbar^{n-1}\pi^{n-1}}\;.
\eeq
where $\mathbf{u}^{\dagger}=(x_0,p_0,x_1,p_1,\cdots,x_{n-1},p_{n-1})$
and $V$, presented in a $(2 \times 2) \otimes (n \times n)$ block form is:
\beq\label{eq:V}
V_{(2n\times 2n)}=\left(%
\begin{array}{llllll}
1 & B& 0& \cdots &0& B^{T}
\\
B^{T} &1 &B & 0& \cdots & 0
\\
0 & B^{T}& 1 & B & 0&\vdots
\\
\vdots & 0 & \ddots & \ddots & \ddots & 0
\\
0 & \cdots & 0&B^{T}&1& B
\\
B & 0 & \cdots &0 & B^{T}& 1
\end{array}
\right)_{(n\times n)}\;, \eeq $1$ and $0$ being the $(2\times 2)$ unit and zero matrices respectively,  and
$B^{T}$ is the transpose of $B=-\frac1{2} \left(%
\begin{array}{rr}
1 & i
\\
-i &1
\end{array}
\right)$.

In order to evaluate the classical limit,
we note that $V$ is a normal matrix and as such it can be written in the form
$V=UDU^{\dagger}$ where $D$  is its diagonal eigenvalue matrix and
$U$ is unitary with orthonormal eigenvector basis as its columns.
Computation of these eigenvectors yields:
\beq \label{eigenV}
\mathbf{e}_{kj}=\frac{1}{\sqrt{2 n}}\left(\begin{array}{c}
(-1)^k\\ i
\end{array}\right)\otimes\left(%
\begin{array}{c}
1\\\om_j\\\om_j^2\\\vdots\\\om_j^{n-1}
\end{array}
\right)_{(1\times n)}\;,
\eeq
with corresponding eigenvalues $\mu_{kj}=1-\om_j^{(-1)^k}$ where $\om_j=e^{2 \pi i j/n}$,
$k=1,2$ and $j=0,...,n-1$. Noting that
$\mu_{1,0}=\mu_{2,0}=0$, the term $\mathbf{u}^\dag V\mathbf{u}$ in the exponent
of (\ref{eq:AnQtoAnC})
can thus be simplified to
\beq\label{eq:decomp}
&&\mathbf{u}^\dag V\mathbf{u}=
\mathbf{v}^\dag D\mathbf{v}
=\sum_{k=1}^2\sum_{j=1}^{n-1}\mu_{kj}v_{kj}^2\;,
\eeq
with $\mathbf{v}^{\dagger} \equiv \mathbf{u}^{\dagger} U$. The limit in (\ref{eq:AnQtoAnC}) thus becomes:
\beq\label{eq:delta}
\lhz \frac{\exp[-\hbar^{-1}\mathbf{u}^\dag V\mathbf{u}]}{\hbar^{n-1}\pi^{n-1}}& = &
\lhz \frac{\exp[-\hbar^{-1}\sum_{k=1}^2\sum_{j=1}^{n-1}\mu_{kj}v_{kj}^2]}{\pi^{n-1}\hbar^{n-1}}
\\\nonumber=\frac{\prod_{k,j}\delta(v_{kj})}{\sqrt{\prod_{k,j}\mu_{kj}}}
&=&\frac{1}{n} \prod_{k,j}\delta( \mathbf{e}_{kj}^{\dagger} \mathbf{u})=
\frac{1}{n}\delta(U^{\dagger}_r\mathbf{u})
\eeq
where we have used the fact that
$\prod_{k,j}\mu_{kj}=\prod_{j=1}^{n-1}(1-e^{2 \pi i j/n})(1-e^{-2 \pi i j/n})=n^2$
and $U^{\dagger}_r$ denotes the conjugate--transpose of the eigenvalue matrix $U$
with its first two eigenvector--columns (corresponding to the zero eigenvalues) removed.
Rewriting $\mathbf{u}$ and $U^{\dagger}_r$ as
\beq
\mathbf{u}^{\dagger}&=&\mathbf{u}_0^{\dagger} \oplus \mathbf{u}_i^{\dagger} \equiv
(x_0,p_0) \oplus (x_1,p_1,x_2,p_2,\cdots,x_{n-1},p_{n-1}) \\\nonumber
U^{\dagger}_r&=& U^{\dagger}_0 \oplus U^{\dagger}_i
\equiv
\frac1{\sqrt{2 n}}
\left(\begin{array}{cc}
-1 & -i \\ 1 & -i
\end{array}\right)
\otimes
\left(
\left( \begin{array}{c}
1 \\ 1\\ \vdots \\ 1
\end{array} \right)
\oplus
\left( \begin{array}{cccc}
\om^{-1}_1 & \om_1^{-2} & \cdots & \om_1^{-(n-1)}
\\
\om^{-1}_2 & \om_2^{-2} & \cdots & \om_2^{-(n-1)}
\\
& & \vdots &
\\
\om^{-1}_{n-1} & \om_{n-1}^{-2} & \cdots & \om_{n-1}^{-(n-1)}
\end{array} \right)
\right)
\eeq
it's easy to show that:
\beq \label{eq:delta3}
\frac{1}{n}\delta(U^{\dagger}_r\mathbf{u})&=&
\frac{1}{n}\delta(U^{\dagger}_i \mathbf{u}_i+U^{\dagger}_0 \mathbf{u}_0) \\\nonumber
&=& \delta(\mathbf{u}_i+U_i U^{\dagger}_0 \mathbf{u}_0)
=\prod_{i=1}^{n-1} \delta(x_i-x_0)\delta(p_i-p_0)\;.
\eeq
Here we have used
$\delta(M \mathbf{x}-\mathbf{n})=|\det M|^{-1} \delta(\mathbf{x}-M^{-1}\mathbf{n})$,
$|\det U_i|=n^{-1}$, and
$(U_i U^{-1}_0 \mathbf{u}_0)^{\dagger}=(x_0,p_0,x_0,p_0,\cdots,x_0,p_0)$.
Using  (\ref{eq:delta3}), we arrive at the final result:
\beq\label{eq:AnQtoAnCFinal}
\lhz \lan A^{n-1}\ran_Q& = & \int \left( \prod_{i=1}^{n-1} dx_i dp_i A(x_i,p_i)
\delta(x_i-x_0)\delta(p_i-p_0)\right)
\\\nonumber =A^{n-1}(x_0,p_0)& \equiv &\lan A^{n-1} \ran_C\;,
\eeq
and so it follows by linearity that (\ref{eq:General}) also holds.
\par
In an exact analogy, it is easy to work out the expectation value
of the multiplication of any two operators $\hat{A}_1$ and $\hat{A}_2$ of the form (\ref{eq:AqOp}):
\beq\label{eq:A1A2QtoC}
\lan \hat{A}_1 \hat{A}_2 \ran_Q&\equiv& Tr [\hat{\rho}_{(q_0,k_0)}\hat{A}_1 \hat{A}_2]\\\nonumber
&=&\int \left( \prod_{i=1}^{2} \pi^{-1} dq_i dk_i A_i(x_i,p_i) \right)
\lan\al_{0}|\al_{1}\ran\lan\al_{1}|\al_{2}\ran
\lan\al_{2}|\al_{0}\ran\;.
\eeq
and verify that in the classical limit it becomes
\beq
\lhz \lan \hat{A}_1 \hat{A}_2 \ran_Q&=&
\int \left( \prod_{i=1}^{2} dx_i dp_i A_i(x_i,p_i)
\delta(x_i-x_0) \delta(p_i-p_0)\right)\\\nonumber
&=&A_1(x_0,p_0) A_2(x_0,p_0)=\lan A_1 A_2 \ran_C
\eeq
Thus, in the classical limit, the expectation value of the
commutator of any two such operators vanishes
\beq \label{comrel}
\lhz \lan [\hat{A}_1 ,\hat{A}_2 ] \ran_Q =\lan A_1 A_2 \ran_C - \lan A_2 A_1 \ran_C =0 \;,
\eeq
as one would expect.
\par
Another important property of the classical--quantum correspondence suggested above is:
\beq \label{comRel2}
\lhz\frac{\lan[\hat{A}_1 ,\hat{A}_2 ]\ran_Q}{\hbar} \; \overset{\rm{l'H\hat{o}pital}}{=} \; \lhz \frac{\partial}{\partial \hbar} \lan[\hat{A}_1 ,\hat{A}_2
]\ran_Q= i \lan \{ A_1 ,A_2 \} \ran_C
\eeq
where $\{A_1,A_2\} \equiv \frac{\partial A_1}{\partial x}
\frac{\partial A_2}{\partial p} -
 \frac{\partial A_1}{\partial p} \frac{\partial A_2}{\partial x}$
stands for the Poisson brackets of the corresponding classical operators. Explicitly written:
\beq\label{A12}
&&\frac{\partial}{\partial \hbar}
\lan[\hat{A}_1 ,\hat{A}_2 ]\ran_Q
\\\nonumber &=& \frac{\partial}{\partial \hbar} \left(\frac1{\pi^2} \int dq_1 dk_1 dq_2 dk_2 \left(
A_1(x_1,p_1)A_2(x_2,p_2) - A_2(x_1,p_1)A_1(x_2,p_2)\right)  \lan\al_{0}|\al_{1}\ran\lan\al_{1}|\al_{2}\ran
\lan\al_{2}|\al_{0}\ran\ \right)\\\nonumber &=& \int dx_1 dp_1 dx_2 dp_2 \left( A_1(x_1,p_1)A_2(x_2,p_2) -
A_2(x_1,p_1)A_1(x_2,p_2)\right) \times \frac{\partial}{\partial \hbar} \left(
\frac{\exp[-\hbar^{-1}\mathbf{u}^{\dag} V\mathbf{u}]}{\hbar^{2}\pi^{2}}\right)
\eeq
where $\mathbf{u}^{\dagger}=(x_0,p_0,x_1,p_1,x_2,p_2)$ and $V$ is
as in (\ref{eq:V}) with $n=3$. Noting that the Gaussian in (\ref{A12}) obeys
\beq
\frac{\partial}{\partial
\hbar}=\frac1{4} \overrightarrow{\nabla}^{\dag}
\left(%
\begin{array}{cc}
1 & -B^{T} \\
-B & 1
\end{array}
\right)
\overrightarrow{\nabla}
\eeq
where $\overrightarrow{\nabla}^{\dag}=(\frac{\partial}{\partial x_1},\frac{\partial}{\partial
p_1},\frac{\partial}{\partial x_2},\frac{\partial}{\partial p_2})$,
we can carefully integrate by parts finally arriving to:
\beq \label{A12Lim}
\lhz \frac{\partial}{\partial \hbar} \lan[\hat{A}_1 ,\hat{A}_2 ]\ran_Q &=& \\\nonumber&=&i \int dx_1 dp_1 dx_2 dp_2 \{
A_1(x_1,p_1),A_2(x_2,p_2)\} \times  \lhz  \frac{\exp[-\hbar^{-1}\mathbf{u}^{\dag}
V\mathbf{u}]}{\hbar^{2}\pi^{2}}\\\nonumber & =&i \lan \{ A_1 ,A_2 \} \ran_C \;. \eeq
\par
So far, we have worked out the classical limit of quantum states, which
correspond to pure classical states (represented by points in phase space).
A generalization of this correspondence may be made
to classical statistical distributions as well. These would be defined by
non--negative functions $P(x,p)$ over phase--space,
with $\int dxdpP(x,p)=1$.
In this case, the classical expectation value for a function $f(A)$ of a classical
observable $A(x,p)$ is given by:
\beq\label{eq:fAc}
\lan f(A) \ran_C=\int dxdp P(x,p) f(A(x,p))\;.
\eeq
The corresponding quantum state assigned to a classical distribution $P(x,p)$
is the following density matrix, given here in a $P$--representation form \cite{Glauber, Sudarshan}:
\beq\label{eq:RhoP}
\hat{\rho}_{P}\equiv \int dxdp P(x,p) \hat{\rho}_{(q,k)}\;.
\eeq
In this case, the quantum expectation value of the $n$th moment
of the quantum observable $\hat{A}$
operating on $\hat{\rho}_{P}$ is given by:
\begin{eqnarray}\label{eq:AQn}
\lan \hat{A}^{n}\ran_Q\equiv Tr [\hat{\rho}_{P}\hat{A}^{n}]=\int dxdp P(x,p)
Tr [\hat{\rho}_{(q,k)}\hat{A}^{n}] \;.
\eeq
Using the  result from (\ref{eq:AnQtoAnCFinal}),
the $\hz$ limit of (\ref{eq:AQn}) simply becomes:
\beq\label{eq:AnQtoAnCRhoP}
\lhz \lan \hat{A}^n\ran_Q=\int dx dp P(x,p) A^n(x,p)\equiv \lan A^n\ran_C \;,
\eeq
and so we can conclude that (\ref{eq:General}) holds for arbitrary classical
distributions \cite{note1} and by the same token, it
is easy to show that the limits of the commutators worked out in (\ref{comrel}) and
(\ref{comRel2}) hold for states of the form (\ref{eq:RhoP}) as well.
\par
A generalization of the scheme given above to
states with many degrees of freedom can be carried out in a straightforward manner
by replacing each phase space point $(x,p)$ with a pair
of vectors $(\mathbf{x},\mathbf{p})$
and each quantum state $\hat{\rho}_{(q,k)}$
with $\hat{\rho}_{(\mathbf{q},\mathbf{k})} \equiv \prod_{\otimes} \hat{\rho}_{(q_i,k_i)}$.
\par
As an example for an immediate application of our proof above,
let us look at the classical
limit of the relative entropy $S(\hat{\rho}_{1} |\hat{\rho}_{2} )$ of two arbitrary
quantum states $\hat{\rho}_{1}$ and $\hat{\rho}_{2}$ constructed
by the classical distributions $P_1$ and $P_2$ respectively, using (\ref{eq:RhoP}).
The relative entropy is defined by
$S(\rho_1 | \rho_2) \equiv \lan \log \hat{\rho}_1 - \log \hat{\rho}_2 \ran_{\hat{\rho}_1}$
\cite{RelDef},
and taking its classical limit, one arrives at
\beq
\lhz S(\hat{\rho}_{1} |\hat{\rho}_{2} ) &=& \lhz \lan \log \hat{\rho}_{1} -
\log \hat{\rho}_{2}\ran_{\hat{\rho}_{1}} \\\nonumber
&=& \lan \log P_1 - \log P_2 \ran_{P_1}
=\int dx dp P_1 ( \log P_1 - \log P_2 ) \equiv \mathcal{K}(P_1 | P_2) \;,
\eeq
which is the relative entropy of the corresponding classical distributions $P_1$ and $P_2$,
also known as the Kullback-Leibler information distance \cite{Kull}.
\par
Up to this point, we have considered the classical limit of a
particular set of quantum states,
but establishing a quantum-classical
correspondence involves equations of motions as well.
In order to ensure that the time evolution
of a quantum system becomes, in the classical limit, that
of its corresponding classical one, let
us now prove that the correspondence still holds
under time--evolution.
This would be done by showing that
\beq \label{TimeEv}
\lhz \frac{d}{dt} \lan \hat{A} \ran_Q = \frac{d}{dt} \lan A \ran_C \;
\eeq
for every classical operator $A$ and its corresponding $\hat{A}$.
To do this, we recall that the equations of motion
for these operators are given by
\beq \label{eqom}
\frac{d}{dt} \lan A \ran_C &=&
 \lan  \{A,H\} \ran_C + \frac{\partial}{\partial t} \lan A \ran_C
\\\nonumber
\frac{d}{dt} \lan \hat{A} \ran_Q &=&-i
\frac{ \lan  [\hat{A},\hat{H}] \ran_Q}{\hbar} + \frac{\partial}{\partial t} \lan \hat{A}\ran_Q
\;,
\eeq
where $H$ is the Hamiltonian governing the time--evolution of
the classical system and $\hat{H}$ is its corresponding quantum one.
Using (\ref{comRel2}), it is easy to verify the right--hand--side
of the quantum equation of motion becomes in the classical limit the right--hand--side
of the classical one, proving (\ref{TimeEv}).\par
To sum up, in this paper
we have presented a scheme which maps classical
states as well as observables to quantum ones, such that
in the $\hz$ limit the latter converges to the former. Moreover, this
correspondence holds under time evolution. We conclude by remarking
that although this mapping is one--to--one it is
certainly not onto; within this scheme there exist quantum states and observables
which cannot be constructed from classical ones.
\par
We thank Nir Lev for his help with the mathematical finer points
and a special thanks to Ady Mann for
his insight and invaluable comments.


\begin{thebibliography}{99}
\bibitem{note1}
It should be noted that
there exist density matrices (\ref{eq:RhoP}) with nonpositive distributions
$P(q,k)$ \cite{Glauber2}, so the
quantum--classical correspondence is not one--to--one,
\textit{i.e.} there are quantum states which do not have classical counterparts.
\end{thebibliography}
\end{document}